\documentclass[aps,showkeys,twocolumn,amsmath,amssymb,superscriptaddress]{revtex4}
\usepackage{graphicx}
\usepackage{dcolumn}
\usepackage{amsmath}
\usepackage{bm}
\usepackage{natbib}

\begin{document}

\title{Measuring spike train synchrony}


\author{Thomas Kreuz}
\email{tkreuz@ucsd.edu} \affiliation{Istituto dei Sistemi Complessi - CNR, Via Madonna del Piano 10, I-50019
Sesto Fiorentino, Italy}
\author{Julie S. Haas}
\affiliation{Institute for Nonlinear Sciences, University of California, San Diego, CA, USA}
\author{Alice Morelli}
\affiliation{Istituto Nazionale di Ottica Applicata, Firenze, Italy}
\author{Henry D. I. Abarbanel}
\affiliation{Institute for Nonlinear Sciences, University of California, San Diego, CA, USA}
\affiliation{Department of Physics and Marine Physical Labaratory (Scripps Institution of Oceanography),
University of California, San Diego, CA, USA}
\author{Antonio Politi}
\affiliation{Istituto dei Sistemi Complessi - CNR, Sesto Fiorentino, Italy}


\date{\today}

\begin{abstract}
Estimating the degree of synchrony or reliability between two or more spike trains is a frequent task in both
experimental and computational neuroscience. In recent years, many different methods have been proposed that
typically compare the timing of spikes on a certain time scale to be optimized by the analyst. Here, we propose
the ISI-distance, a simple complementary approach that extracts information from the interspike intervals by
evaluating the ratio of the instantaneous firing rates. The method is parameter free, time scale independent and
easy to visualize as illustrated by an application to real neuronal spike trains obtained in vitro from rat
slices. In a comparison with existing approaches on spike trains extracted from a simulated Hindemarsh-Rose
network, the ISI-distance performs as well as the best time-scale-optimized measure based on spike timing.
\end{abstract}


\newcommand{\abb}{\small\sf}

\maketitle

%
%
\section{\label{s:Intro} Introduction}

The basic elements of neuronal communication are pulsed electric signals called action potentials or spikes.
Under the assumption that both the shape of the spike and the background activity carry minimal information,
neuronal responses are typically reduced to the much simpler form of a spike train, where the only information
maintained is the timing of the single spikes. Measuring the overall degree of synchrony between different spike
trains is an important tool in many different contexts. It can be used to quantify the reliability of responses
upon repeated presentations of the same stimulus (\citeauthor{Mainen95}, \citeyear{Mainen95}), to address
questions regarding the limitations of neuronal coding (rate versus time coding) (\citeauthor{Rieke96},
\citeyear{Rieke96}) or to evaluate the information transfer between synaptically coupled neurons (cf., e.g.,
\citeauthor{Reyes03}, \citeyear{Reyes03}).

A variety of different measures have been introduced to address the synchrony between spike trains. Most of
these measures require considering a large number of trials, not just two. Some of them are based on the
construction of a post-stimulus time histogram (PSTH), derived from multiple trials. PSTH measures such as
reliability, precision and sparseness (\citeauthor{Mainen95}, \citeyear{Mainen95}; \citeauthor{Berry97},
\citeyear{Berry97}) rely on the analyst to define the so-called events, i.e., bursts of high firing rate. Other
methods (i) quantify the occurrence of given spike patterns and measure their robustness (``attractor
reliability", \citeauthor{Tiesinga02b}, \citeyear{Tiesinga02b}), (ii) exploit the deviation of the spike train
statistics from a Poissonian distribution (\citeauthor{Brenner00}, \citeyear{Brenner00};
\citeauthor{Tiesinga04}, \citeyear{Tiesinga04}), or (iii) measure the normalized variance of pooled
exponentially convolved spike trains (\citeauthor{Hunter98}, \citeyear{Hunter98}).

The focus of this study lies on a group of measures that aim at a quantification of the degree of similarity or
dissimilarity between as few as two spike trains. A very prominent example of such bivariate approaches are
spike train distances that consider spike trains to be points in an abstract metric space and assign
non-negative values quantifying the dissimilarity between a given pair of spike trains. Among these is the
distance introduced in \citeauthor{Victor96} (\citeyear{Victor96}), which evaluates the ``cost" needed to
transform one spike train into the other, using only certain elementary steps. Another metric proposed in
\citeauthor{VanRossum01} (\citeyear{VanRossum01}), measures the Euclidean distance between the two spike trains
after convolution of the spikes with an exponential function. Other approaches quantify the cross correlation of
spike trains after exponential or Gaussian filtering (\citeauthor{Haas02}, \citeyear{Haas02};
\citeauthor{S_Schreiber03}, \citeyear{S_Schreiber03}), or exploit the exponentially weighted distance to the
nearest neighbor (\citeauthor{Hunter03}, \citeyear{Hunter03}). A common property of all these measures is the
existence of one parameter that sets the time scale for the analysis. This introduces elements of human
fallibility and presumptions into the analysis and furthermore discourages direct comparisons between different
sets of results. Such a parameter does not exist for event synchronization, a method proposed in
\citeauthor{QuianQuiroga02b} (\citeyear{QuianQuiroga02b}) that quantifies the number of quasi-simultaneous
appearances, using a variable time scale that automatically adapts itself to the local spike rates.

In this study a measure is proposed that, complementary to the approaches mentioned above, uses the interspike
interval (ISI) instead of the spike as the basic element of comparison. The ISI-distance quantifies the ratio of
instantaneous rates and facilitates visualization of the relative timing of pairs of spike trains. Since no
binning is used, the measure has the maximum possible time resolution (i.e., up to a single spike). Similar to
event synchronization, it is both parameter--free and self-adaptive so that there is no need to fix a time scale
beforehand. Thus the analyst is removed from the analysis, allowing for a more objective and broadly comparable
measure of neuronal synchronization. In the first part of this study, the ISI-distance is illustrated using real
in vitro data from cortical cells.

Moreover, since a comparison between different approaches was still missing, in the second part we tested the
performance of several bivariate measures, including the ISI-disance, using clustered spike trains in a
controlled setting. These were generated from a network of simulated Hindemarsh-Rose neurons with a
pre-determined degree of coupling between pairs. In this scenario, different spike trains belonged to different
clusters and the capability of the measures to detect the original clustering behavior could be quantified by
two indices, which evaluate the correctness of the clusters and the separation between them, respectively.
However, we do not claim to assess the absolute validity of our or any other measure, as no single number can
represent the true code of the spike-generating mechanism under any circumstances. In a final step, to assess
the similarity of the different approaches to measure spike train synchrony, we evaluated the degree of
redundance between the different measures by means of a correlation analysis \footnote{In a complementary study
(\citeauthor{Kreuz07a}, \citeyear{Kreuz07a}) a similar performance comparison and correlation analysis has been
carried out for bivariate measures that quantify the synchrony between continuous time series (and not just
discrete events such as spikes).}.

The remainder of the paper is organized as follows: In the methods section, after a short description of the
spike detection algorithm (Section \ref{ss:Methods-Spike-Detection}), a more detailed description of the new
method based on the ISI-distance is given (Section \ref{ss:Methods-ISI-Distance}). It is illustrated using in
vitro recordings from cortical cells in the entorhinal cortex of rats. The following section,
\ref{ss:Methods-Measures}, contains a short overview over the existing measures against which this new method is
compared. The cluster analysis is described in Section \ref{ss:Methods-Clustering}, while the correlation
analysis is described in section \ref{ss:Methods-Correlation}. In section \ref{ss:Results-Comparison} the actual
comparison of the different methods on simulated time series taken from a network of Hindemarsh-Rose
model-neurons is carried out. Conclusions are drawn in section \ref{s:Discussion}.  Finally, both data sets (the
in vitro recordings and the simulated Hindemarsh-Rose time series) are described in the appendix in sections
\ref{ss:Data-Stellate-Cells} and \ref{ss:Data-Hindemarsh-Rose}, respectively.

%
%
\section{\label{s:Methods} Methods}

\subsection{\label{ss:Methods-Spike-Detection} Spike detection}

A prerequisite to any method is the extraction of the spike times from the time series by means of a standard
spike detection algorithm. Typically, some sort of threshold criterion is employed, either for the time series
itself or its derivative. Thereby the continuous time series is transformed into a discrete series of spikes.
Each spike train can then be expressed as a series of $\delta$ functions
\begin{equation} \label{eq:Spike-Train}
    S (t) = \sum_{i=1}^{M} \delta(t - t_i)
\end{equation}
with ${t_1,...t_M}$ denoting the series of spike times and $M$ being the number of spikes.

In this study, for all the different measures the same spike detection algorithm is used. The threshold is
chosen as the arithmetic average over the minimum and maximum value of the action potential.

\subsection{\label{ss:Methods-ISI-Distance} The ISI-distance}

\begin{figure}
    \includegraphics[width=85mm]{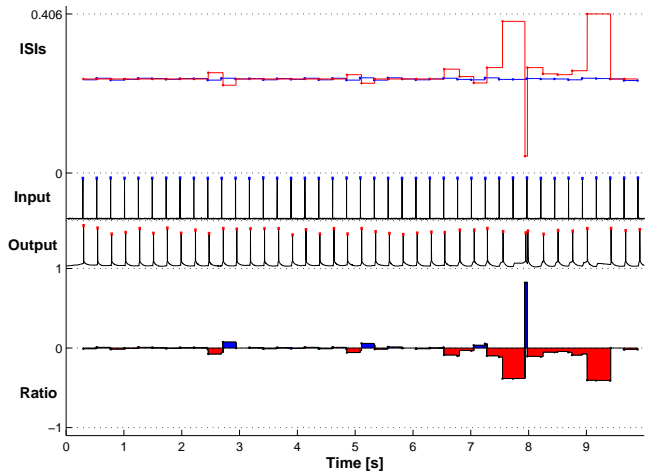}
    \caption{\abb\label{fig:ISI-Distance-RealData-1} (color online) First example of cortical cell recordings. In the middle traces the
        two recorded time series are shown. The detected spikes are marked in blue (input) and red (output),
        respectively. On top the ISI-values according to Eq. \ref{eq:ISI} are depicted, at the bottom the
        corresponding renormalized ISI-distance (cf. Eq. \ref{eq:ISI-Distance}). Here colors mark the times where the
        respective spike train is slower. For this pair of spike trains an ISI-distance $D_I = 0.06$ is obtained.}
\end{figure}

\begin{figure}
    \includegraphics[width=85mm]{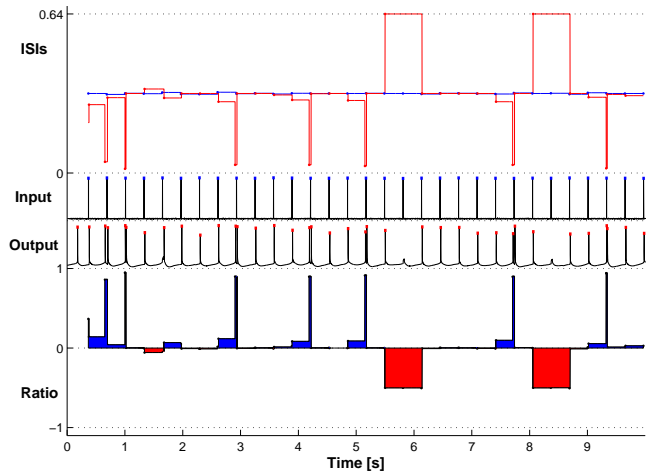}
    \caption{\abb\label{fig:ISI-Distance-RealData-2} (color online) Second example of cortical cell recording shown in the same way as in
        Fig. \ref{fig:ISI-Distance-RealData-1}. In this case the ISI-distance attains the value $D_I = 0.11$.}
\end{figure}

\begin{figure}
    \includegraphics[width=85mm]{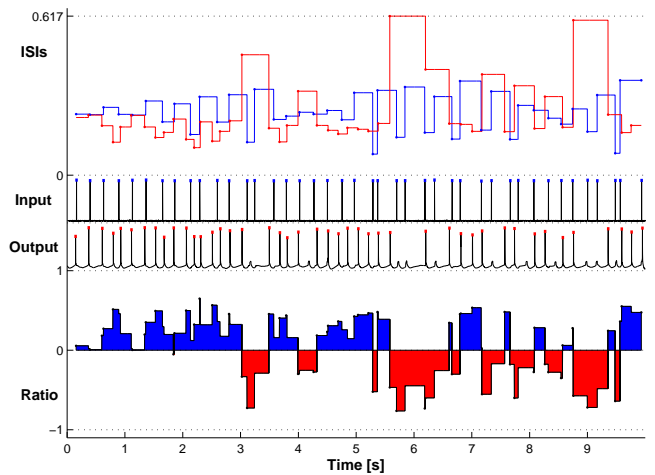}
    \caption{\abb\label{fig:ISI-Distance-RealData-3} (color online) Third example of cortical cell recording shown in the same way as in
        Fig. \ref{fig:ISI-Distance-RealData-1}. The ISI-distance for this example is $D_I = 0.34$.}
\end{figure}

To obtain a time-resolved measure of the firing rate of the spike train $\{t^x_i\}$, in a first step the value
of the current interspike interval is assigned to each time instant \footnote{This is closely related to the
Rice phase, that is obtained by linear interpolation between two events (e.g., spikes) from zero to $2\pi$ (cf.
e.g. \citeauthor{Callenbach02}, \citeyear{Callenbach02}; \citeauthor{Freund03}, \citeyear{Freund03}). In fact,
the measure $I$ is proportional to the ratio of instantaneous Rice frequencies, however, the normalization used
here allows for a better visualization.}(see
Figs.~\ref{fig:ISI-Distance-RealData-1}-\ref{fig:ISI-Distance-RealData-3}, top),
\begin{equation} \label{eq:ISI}
    x_{isi} (t) = \min(t^x_i | t^x_i > t) - \max(t^x_i | t^x_i < t)
     \quad  t^x_1 < t < t^x_{M}
\end{equation}
and accordingly for the second spike train $\{t^y_j\}$. Now, in a second step the ratio between $x_{isi}$ and
$y_{isi}$ is taken (effectively, this is done just once after every new spike in either time series), and the
final measure is thereby obtained after introducing a suitable normalization,
%
%
%

\begin{equation} \label{eq:ISI-distance}
    I (t) = \begin{cases}
           x_{isi} (t) / y_{isi} (t) - 1 & {\rm if} ~~ x_{isi} (t)<= y_{isi} (t) \cr
                      - (y_{isi} (t) / x_{isi} (t) -1)     & {\rm else}.
                  \end{cases}
\end{equation}
The measure becomes zero in case of iso-frequent behavior, and approaches $-1$ and $1$, respectively, if the
firing rate of the first (or second) train is infinitely high and the other infinitely low (see
Figs.~\ref{fig:ISI-Distance-RealData-1}-\ref{fig:ISI-Distance-RealData-3}, bottom).

Finally, in order to derive a measure of spike train distance, there are two possible ways of averaging. In the
time-weighted variant, the absolute ISI-distance is integrated over time,
\begin{equation} \label{eq:ISI-Distance}
    D_I =\int_{t=0}^T dt | I(t) |,
\end{equation}
whereas in the spike-weighted variant, the ISI-distance is evaluated only after every new spike in either time
series,
\begin{equation} \label{eq:ISI-Distance-spike weighted}
    D_I^s =\sum_{i=1}^M | I(t_i) |.
\end{equation}
There are a number of meaningful extensions of this measure. Omitting the absolute values yields a quantity that
evaluates the relative firing rates of the two spike trains. Quantifying higher moments of the
$I(t)$-distribution such as the standard deviation can provide additional information (in particular it allows
to distinguish the cases of random jitter and systematic phase lag that could lead to the same ISI-distance).
Also all of these variants can be implemented using a moving-window analysis. Due to the self-adaptation of this
measure to the time scale of the spikes, reasonable results can be obtained also for rather short spike trains.
Finally, the sensitivity of the measure can be extended to longer time scales and a more coarse-grained
evaluation by averaging $x_{isi} (t)$ and $y_{isi} (t)$ over neighboring ISIs. An example of a possible
application is the quantification of (dis-)similarities between bursts in time series.

In Fig.~\ref{fig:ISI-Distance-RealData-1} the ISI-distance is applied to two exemplary input-output spike trains
of $10 \, s$ duration (for a description of the data see Appendix \ref{ss:Data-Stellate-Cells}). In the first
seconds, the spike trains are $1:1$ synchronized and this is reflected by an ISI-distance $I(t) \approx 0$.
Nevertheless, small deviations can be visualized that are hard to catch from a visual inspection of the spike
trains themselves. These deviations are more pronounced in the second half of the recording where the output no
longer follows the input but rather slows down (as reflected by predominantly negative values marked in red) and
a spike doublet occurs (as indicated by the short excursion to positive values marked in blue). The example
shown in Fig.~\ref{fig:ISI-Distance-RealData-2} reveals that certain patterns in the ISI-distance appear
repeatedly. The output exhibits several spike doublets, some of which are followed by a miss (reflected by the
negative values marked in red). Finally, a more irregular behavior is shown in
Fig.~\ref{fig:ISI-Distance-RealData-3} where it is again clear that the ISI-distance allows tracing the relative
firing rate behavior in a simple way \footnote{The Matlab source code for calculating the ISI-distance and
plotting this kind of figures as well as information about the implementation can be found under
http://inls.ucsd.edu/\textasciitilde kreuz/Source-Code/Spike-Sync.html.}.

\subsection{\label{ss:Methods-Measures} Existing measures of spike train distance}

In this study the ISI-distance will be compared against five existing measures of spike train (dis-)similarity.
These comprise the spike train metrics introduced in \citeauthor{Victor96} (\citeyear{Victor96}), as well as in
\citeauthor{VanRossum01} (\citeyear{VanRossum01}), a correlation measure proposed in \citeauthor{S_Schreiber03}
(\citeyear{S_Schreiber03}), another distance measure introduced in \citeauthor{Hunter03} (\citeyear{Hunter03}),
and event synchronization, a method introduced in \citeauthor{QuianQuiroga02b} (\citeyear{QuianQuiroga02b})
\footnote{We excluded approaches that rely on binning (such as \citeauthor{Johnson01} (\citeyear{Johnson01}),
\citeauthor{Christen06} (\citeyear{Christen06})) from the analysis.}.

In order to compare the various measures, we turned each of them into a suitably-normalized measure of
dissimilarity, as this is more akin to the concept of distance.

\subsubsection{\label{sss:Methods-Victor} Victor-Purpura spike train metric}

The spike train metric $D_V$ introduced in \citeauthor{Victor96} (\citeyear{Victor96}) defines the distance
between two spike trains in terms of the minimum cost of transforming one spike train into the other by using
just three basic operations: spike insertion, spike deletion and spike movement. While the cost of insertion or
deletion of a spike is set to one, the cost $c_V$ of moving a spike is the only parameter of the method setting
the time scale of the analysis. For small $c_V$, the distance basically equals the difference in spike number,
whereas for high $c_V$, the distance approaches the number of non-coincident spikes, since instead of shifting
spikes it becomes more favorable to delete all non-coincident spikes of the one time series and to insert all
non-coincident spikes of the other. Thus, by increasing the cost $c_V$, the distance is transformed from a rate
distance to a timing distance.
%

\subsubsection{\label{sss:Methods-vanRossum} Van Rossum spike train metric}

A second spike train metric was introduced in \citeauthor{VanRossum01} (\citeyear{VanRossum01}). In this method,
each spike is convolved with an exponential function ${\rm e}^{-(t-t_i)/\tau_R}$ ($t>t_{i}$), where $t_i$ is the
spike time. From the convolved waveforms $f(t)$ and $g(t)$, the van Rossum distance $D_R$ can be calculated as
%
%
\begin{equation} \label{eq:RossumDistance}
    D_R (\tau_R) = \frac{1}{\tau_R} \int_0^\infty [ f(t) - g(t) ]^2 dt
\end{equation}
Since the post-synaptic currents triggered by the single spikes approximate exponentials, the van Rossum
distance estimates the difference in the effect of the two trains on the respective synapses. In this method,
the time constant $\tau_R$ of the exponential is the parameter setting the time scale. It is the inverse of
Victor and Purpura's cost parameter: $\tau_R = 1/c_V$.

\subsubsection{\label{sss:Methods-Schreiber} Schreiber et al. similarity measure}

The correlation-based approach was first described in \citeauthor{Haas02} (\citeyear{Haas02}), and later
detailed in \citeauthor{S_Schreiber03} (\citeyear{S_Schreiber03}). In this approach, each spike train is
convolved with a filter of a certain width (exponential in \citeauthor{Haas02} (\citeyear{Haas02}), Gaussian in
\citeauthor{S_Schreiber03} (\citeyear{S_Schreiber03})) to form $s_i$ before cross correlation and normalization.
\begin{equation} \label{eq:SchreiberCorrelation}
    S_S(\sigma_S) = \frac{s_i s_j}{|s_i||s_j|}.
\end{equation}
Haas and White allowed a minimal phase lag in the cross correlation (and thus another parameter to adjust),
while Schreiber et al. allowed none. Here, the approach by Schreiber et al. is used. The width of the convolving
filter $\sigma_S$ sets the time scale of interaction between the two spike trains. The inversion $D_S=1-S_S$
yields a normalized measure of spike train dissimilarity that can be compared with the other. A clustering
analysis based on this measure was performed in \citeauthor{Fellous04} (\citeyear{Fellous04}).

\subsubsection{\label{sss:Methods-Hunter} Hunter-Milton similarity measure}

This approach, introduced in \citeauthor{Hunter03} (\citeyear{Hunter03}), starts by identifying in the spike
train $y$ the nearest spike $t_{k(i)}^y$ to the spike occurring at time $t_i^x$ in the spike train $x$. The
degree of coincidence between these spikes is quantified by
$r_{xy}=\exp{\left(-|t_i^x-t_{k(i)}^y|/\tau_H\right)}$ and the overall similarity measure $S_H$ is thereby
determined as the symmetrized average of $r_{xy}$ over the entire series,
\begin{equation} \label{eq:HunterDistance}
    S_H = \frac{ \langle r_{xy}\rangle + \langle r_{yx}\rangle}{2}.
\end{equation}
Also in this method, there is a free time scale that can be set by fixing the parameter $\tau_H$. For two
identical spike trains $r_{xy}=r_{yx}=1$. Accordingly, a measure of spike train dissimilarity can be obtained as
$D_H=1-S_H$

\subsubsection{\label{sss:Methods-Event-Synchronization} Event synchronization}

The last measure is a variant of the event synchronization proposed in \citeauthor{QuianQuiroga02b}
(\citeyear{QuianQuiroga02b}), and later used in \citeauthor{Hahnloser02} (\citeyear{Hahnloser02}),
\citeauthor{Kreuz04} (\citeyear{Kreuz04}), and \citeauthor{Kreuz07a} (\citeyear{Kreuz07a}). This measure
quantifies the overall level of synchronicity from the number of quasi-simultaneous appearances of spikes.
However, in contrast to the measures introduced above, this method is scale-free, since the maximum time lag
$\tau_{ij}$ up to which two spikes $t_i^x$ and $t_j^y$ are considered to be synchronous is adapted to the local
spike rates according to
\begin{equation} \label{eq:Event-MaxDist}
    \tau_{ij} = \min \{t_{i+1}^x - t_i^x, t_i^x - t_{i-1}^x,
                           t_{j+1}^y - t_j^y, t_j^y - t_{j-1}^y\}/2.
\end{equation}
Then the function $c(x|y)$ is introduced to count the number of times a spike appears in $x$ shortly after a
spike appears in $y$,
\begin{equation} \label{eq:Event-Count}
    c (x|y) = \sum_{i=1}^{M_x} \sum_{j=1}^{M_y} J_{ij},
\end{equation}
where
\begin{equation} \label{eq:Event-Synchronicity}
    J_{ij} = \begin{cases}
                      1     & {\rm if} ~~ 0 < t_i^x - t_j^y \leq \tau_{ij} \cr
                      1/2   & {\rm if} ~~ t^x_i = t^y_j \cr
                      0     & {\rm else}.
                  \end{cases}
\end{equation}
With $c(y|x)$ defined accordingly, we can write the event synchronization as
\begin{equation} \label{eq:Event-Synchro}
    Q = \frac {c (y|x) + c (x|y)} {\sqrt{M_x M_y}}.
\end{equation}
Again, a measure of spike train distance can be defined as $D_Q=Q-1$. With the above normalization, $0 \leq D_Q
\leq 1$, with $D_Q = 0$ if and only if all spikes of the signals are synchronous \footnote{This normalization is
superior to the normalization proposed for a similar measure in \citeauthor{Tiesinga04b}
(\citeyear{Tiesinga04b}). There, the so called coincidence factor was normalized to the minimum number of spikes
in either spike train. With that normalization it would, in the extreme case, be possible that a single spike
can synchronize perfectly with a very long spike train just because it coincides (maybe by chance) with one of
the spikes in the other sequence. On the other hand, with the normalization used here, the maximum value can
only be achieved for truly synchronous spike trains, i.e., a difference in spike number is correctly reflected
as a first deviation from perfect synchrony and a $Q$ value lower than one is assigned. As an additional
confirmation, we also find that the clustering performance is superior for the normalization proposed here
(results not shown).}.

\subsection{\label{ss:Methods-Clustering} Assessing clustering quality}

One important application for measures of spike train synchrony is the identification of interspike correlations
and the reconstruction of clustering patterns. In order to test the above measures, we generated 29 spike trains
by simulating a network of Hindmarsh-Rose neurons (see appendix \ref{ss:Data-Hindemarsh-Rose}). From the network
architecture and the pattern coding, we organized the $29$ spike trains into three principal clusters with $13$
members (clusters $1$ and $2$) and $3$ members (cluster $S$), respectively. We first validated the different
measures and then quantified their performance in reproducing the cluster structure by means of two indices. For
the four measures $D_V$, $D_R$, $D_S$ and $D_H$ that depend on a parameter setting the time scale, we varied the
respective parameter over several logarithmic decades, with four equidistant values within each decade, i.e.,
$c_V = 10^{-a+0.25 b}$. In each case, we adapted the parameter range via $a$ and $b$ to cover the relevant
extreme cases.

\begin{figure}
    \includegraphics[width=85mm]{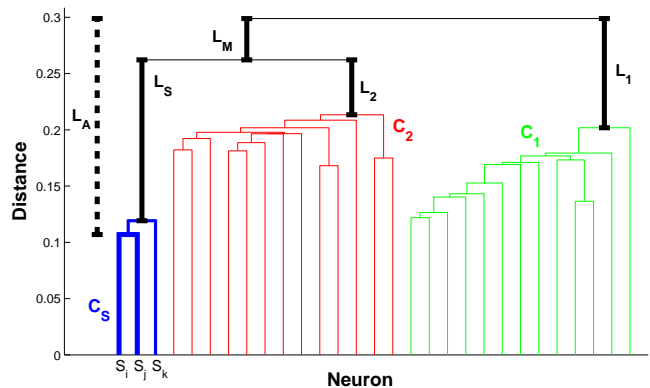}
    \caption{\abb\label{fig:Cluster-Example} (color online) Example of a hierarchical cluster tree obtained from $29$ Hindemarsh-Rose
        time series employing the event distance $D_Q$. The three principal clusters $C_1$, $C_2$ and $C_S$ are
        distinguished by different colors. The merging of the first two time series $S_i$ and $S_j$ to cluster
        $C_\alpha$ is highlighted by very thick blue lines, the consecutive merging of this cluster with $S_k$ by
        thick blue lines. Finally, black lines mark the separation of the three principal clusters used for the
        definition of the cluster-separation $F$. The clustering performance values for this example are $H = 1$ and
        $F = 0.57$.}
\end{figure}

After applying a given similarity measure to all possible pairs of spike trains, we generated a hierarchical
cluster tree (dendrogram) by applying the single linkage algorithm provided by Matlab to the resulting pairwise
distance matrices. An exemplary dendrogram obtained from the $29$ Hindemarsh-Rose time series using the event
distance $D_Q$ is shown in Fig.~\ref{fig:Cluster-Example}. Three principal clusters are clearly distinguished.
The dendrogram is constructed as follows: First, the closest pair $S_i$, $S_j$ of sequences is identified and
thereby linked by a $\sqcap$-shaped line, where the height of the connection measures the mutual distance
$d(S_i,S_j)$. These two time series are merged into a single element $C_\alpha$, and the next closest pair of
elements is then identified and connected. The procedure is repeated iteratively until a single cluster remains.
The implementation of the method requires introducing the distance between a pair of clusters $C_\alpha$,
$C_\beta$. In the single linkage algorithm, it is defined as the minimum over all the distances between pairs of
sequences in the two clusters, i.e., $d(C_\alpha,C_\beta)=\min\{d(S_k,S_m)\}$, $S_k\in C_\alpha$, $S_m\in
C_\beta$.

In order to quantify the success in reproducing this clustering, we computed the entropy of the confusion matrix
$N_{\alpha \beta}$) (\citeauthor{Abramson63}, \citeyear{Abramson63}; \citeauthor{Victor96},
\citeyear{Victor96}). The entry $N_{\alpha\beta}$ is defined as the number of times $S_\beta$ is the closest
cluster to a spike train belonging to $S_\alpha$. Following \citeauthor{Victor96}, the distance between the
spike train $S_i$ and the cluster $C_\alpha$ is defined as $\langle d(S_i,S_j) \rangle_\alpha$, where $\langle
\cdot \rangle_j$ denotes the average over all spike trains in the cluster $C_\alpha$. Note that this distance is
different from the one used for the cluster identification, however, we verified that results proved to be
robust against variations of the distance used. For a perfect clustering $N$ is diagonal, whereas each
misclassification yields a non-diagonal elements. The relative amount of misclassifications is finally
quantified by the entropy
\begin{equation}\label{eq:Entropy}
 H_C = \sum_{\alpha,\beta} p_{\alpha\beta} \log \frac{p_{\alpha\beta}}{P_\alpha Q_\beta}
\end{equation}
where $p_{\alpha \beta} = N_{\alpha \beta}/N_{tot}$, $P_\alpha = \sum_b p_{\alpha b}$, and $Q_\beta = \sum_b
p_{b\beta}$. This entropy value is then normalized to the maximum entropy obtained for a correct classification
\footnote{The bias correction introduced in \citeauthor{Aronov03} (\citeyear{Aronov03}) is omitted since it is
not necessary for a relative comparison of measures.},
\begin{equation} \label{eq:Norm-Entropy}
    H = H_C/H_{max}.
\end{equation}
Although the clustering entropy $H$ evaluates the correctness of the hierarchical tree, it does not quantify the
separation of the three principal clusters in those cases where the expected clustering is obtained. As can be
seen in Fig.~\ref{fig:Cluster-Example}, the cluster separation is given by the lengths $L_1$, $L_2$, $L_S$, and
$L_M$ of their upper branches. A convenient way of quantifying the cluster separation with a single indicator is
by introducing the index
\begin{equation} \label{eq:Cluster-Separation}
    F = \frac{L_1 + L_2 + L_S + L_M}{3 L_A} ,
\end{equation}
where the branch lengths are normalized to the difference $L_A$ between the overall maximum and the overall
minimum distance thus guaranteeing that the $F$-values range in the interval $[0,1]$. Given a correct clustering
with three principal clusters, all quantities needed for the calculation of $F$ can be extracted from the output
matrix of the Matlab single linkage algorithm, otherwise the clustering separation is set to $F = 0$.

\subsection{\label{ss:Methods-Correlation} Correlations between the different measures}

In order to investigate to which extent the different measures of spike train distance carry independent and
non-redundant information, we performed a correlation analysis. First, for each of the four measures $D_V$,
$D_R$, $D_S$, and $D_H$ that depend on a parameter setting the time scale, we identified the parameter value
yielding the best clustering results. Then, we applied each of the six measures ($D_I$, $D_V$, $D_R$, $D_S$,
$D_H$, and $D_Q$) to the different pairs of sequences, obtaining six sets of $29 \times 28/2 = 406$ different
values. In order to guarantee maximal homogeneity, the various measures were all scaled to the $[0, 1]$ range
(this means that the two unnormalized measures $D_V$ and $D_R$ were divided by their maximal values). Moreover,
following a customary approach to better fit a normal distribution, each data set was arcsin-transformed using
$x' = \arcsin(\sqrt{x})$ (\citeauthor{Daniels47}, \citeyear{Daniels47}). Finally, we determined the Pearson
correlation coefficients (\citeauthor{Devore05}, \citeyear{Devore05}) and from the pairwise distances
(1-correlation) among the different measures, we obtained a hierarchical cluster tree (using again the
single-linkage algorithm).

%
%
\section{\label{s:Results} Results}

\subsection{\label{ss:Results-Comparison} Comparison of measures using simulated Hindemarsh-Rose time series}

\begin{figure}
    \includegraphics[width=85mm]{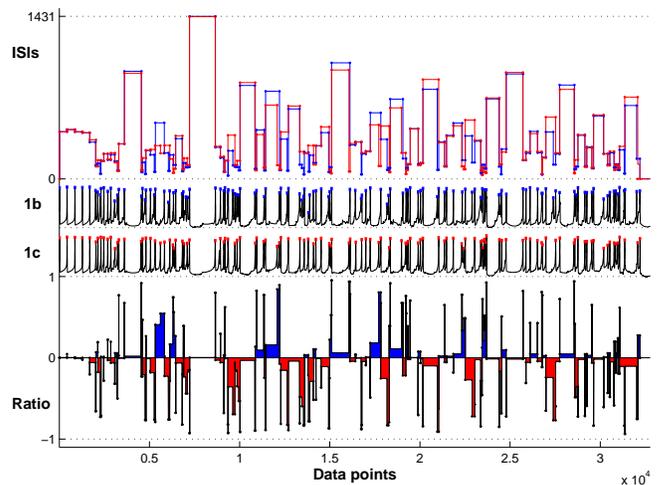}
    \caption{\abb\label{fig:ISI-Distance-1b-1c-Same} (color online) Two time series from two neurons coding both for the
        same pattern B (middle). The detected spikes are marked in blue and red, respectively. On top the
        ISIs, at the bottom the corresponding renormalized ISI-distance. Here colors mark the times where
        the respective spike train is slower. For this combination an ISI-distance $D_I=0.019$ is obtained.}
\end{figure}

\begin{figure}
    \includegraphics[width=85mm]{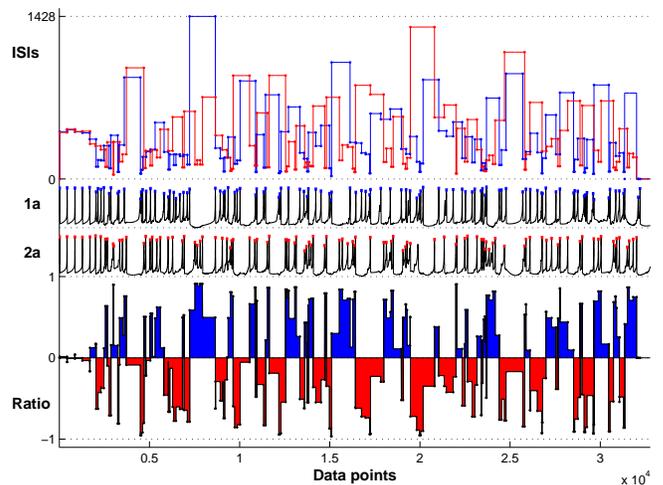}
    \caption{\abb\label{fig:ISI-Distance-1a-2a-Diff} (color online) Same as Fig. \ref{fig:ISI-Distance-1b-1c-Same} but this time for two
        time series from two neurons coding for different patterns. In this case the ISI-distance $D_I=0.032$ is much
        higher.}
\end{figure}

In order to compare the various dissimilarity measures, we have analyzed numerically generated spike trains,
since their properties are much more controllable. More precisely, we refer to Hindemarsh-Rose time series whose
clustering properties are known beforehand (see appendix \ref{ss:Data-Hindemarsh-Rose} for a description of the
underlying model). Two instances of the ISI-distance are shown in Figs.~\ref{fig:ISI-Distance-1b-1c-Same} and
\ref{fig:ISI-Distance-1a-2a-Diff} where the signals emitted by two neurons belonging to the same and to
different clusters, respectively, are compared. In the first example, deviations from zero of the ISI-distance
are confined to short time scales (they are mostly due to small phase shifts that accompany large changes of the
firing rate). In the second example, long-lasting differences are detected which also exhibit a sort of
oscillation.

\begin{figure}
    \includegraphics[width=85mm]{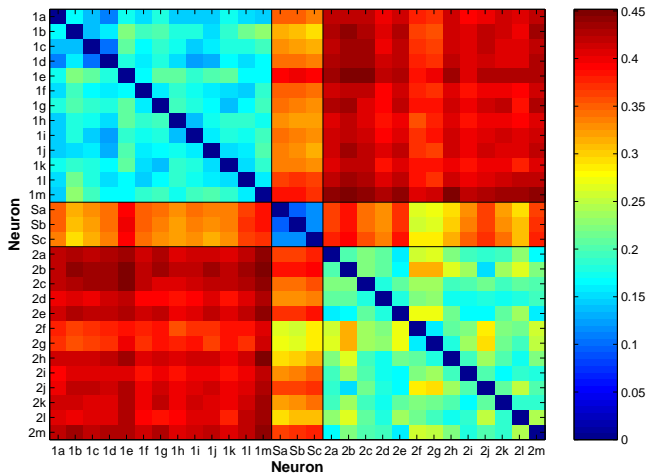}
    \caption{\abb\label{fig:Corr-IR} (color online) Distance matrix for $29$ time series from the Hindemarsh-Rose model.
        Results are obtained by using the ISI-distance $D_I$. Neurons are labelled by '1', '2' and 'S',
        respectively, depending on their affiliation to pattern $1$, $2$ or both ('shared').}
\end{figure}

The distance matrix obtained from the application of the ISI-distance $D_I$ to all $406$ combinations of the
$29$ spike trains is shown in Fig. \ref{fig:Corr-IR}. Patterns can be clearly recognized, since the neurons have
been ordered according to to their a priori affiliation known from the model setup. Smallest distances are
obtained for pairs of spike trains belonging to the same cluster starting from those within $S$. At the other
extreme, the largest distances are found for spike trains belonging to the clusters $1$ and $2$.
\begin{figure}
    \includegraphics[width=85mm]{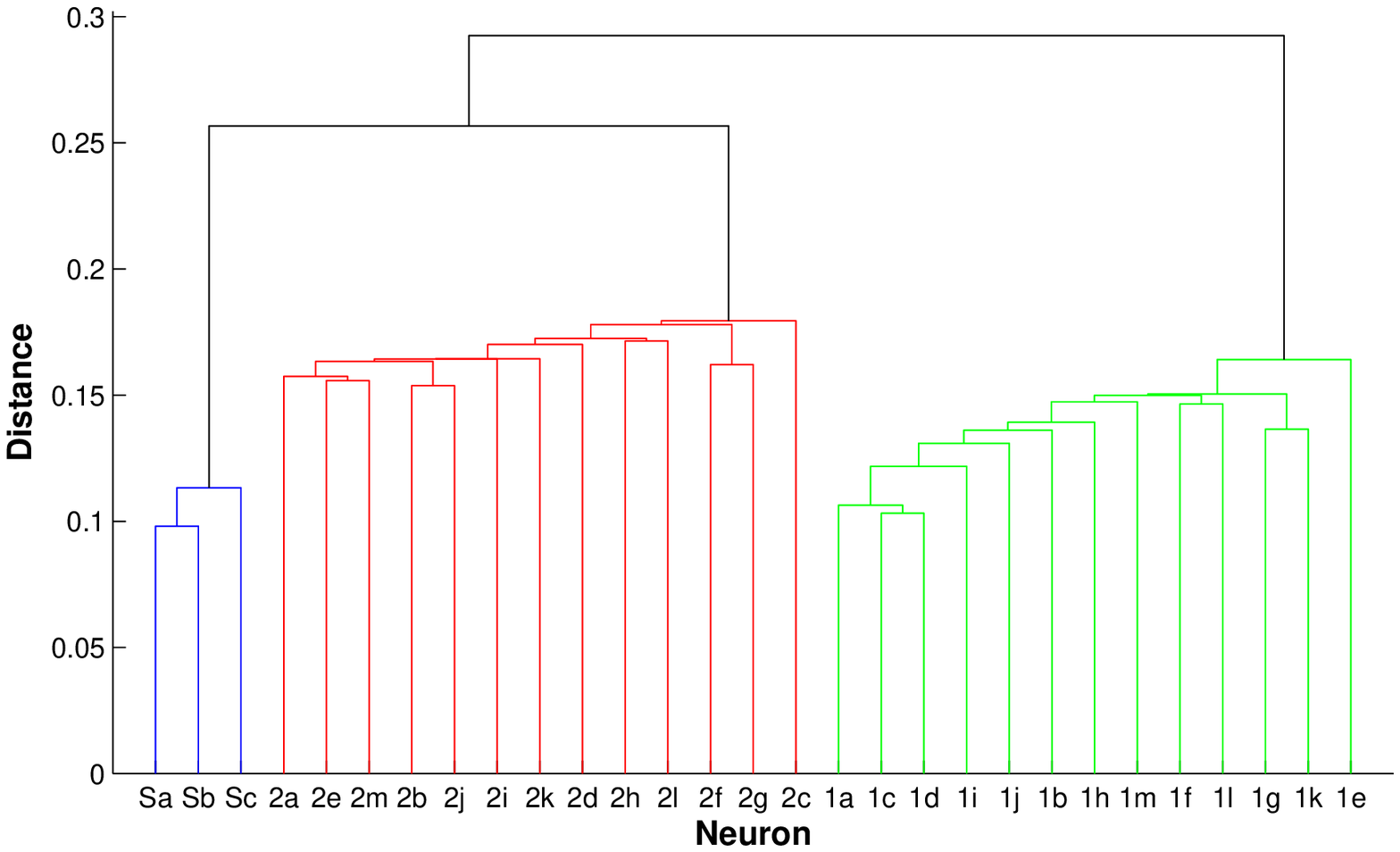}
    \caption{\abb\label{fig:Cluster-IR} (color online) Clustering of $29$ time series from the Hindemarsh-Rose model using the ISI-distance.
        The existence of three clearly separated clusters is evident. The cluster performance values are $H = 1$
        and $F = 0.66$.}
\end{figure}

From this distance matrix, we generated the cluster dendrogram shown in Fig.~\ref{fig:Cluster-IR}, where the
three principal clusters are clearly visible. The absence of misclassifications implies that the clustering
entropy (computed according to Eqs.~(\ref{eq:Entropy},\ref{eq:Norm-Entropy})) is $H = 1$. On the other hand, the
cluster separation determined from Eq.~(\ref{eq:Cluster-Separation}) is for this case $F=0.66$. Similar results
have been obtained for the other parameter-free measure, the event distance $D_Q$ (the corresponding dendrogram
being shown in Fig. \ref{fig:Cluster-Example}). Also in this case, $H = 1$, while the smaller value of $F$
(0.60) suggests a slightly lower clustering quality.

\begin{figure}
    \includegraphics[width=85mm]{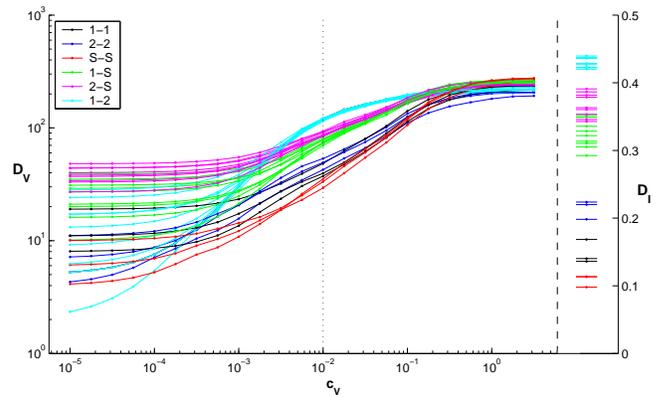}
    \caption{\abb\label{fig:Para-VSD} (color online) Dependence of the Victor-Purpura spike train distance $D_V$ on
        the cost parameter $c_V$. For each combination of cluster affiliations a different color is used. The dotted
        line marks the cost value for which the best clustering separation is obtained. On the right hand side of the
        dashed line the values obtained for the ISI-distance without any optimization are depicted.
        Note that due to different normalizations and scalings, the measures appear on different y-axes.}
\end{figure}
We then investigated the performance of the Victor-Purpura and van Rossum spike train distances, as well as of
the Schreiber et al. and Hunter-Milton dissimilarities for different values of the free parameter in order to
select the proper time scale. In Fig.~\ref{fig:Para-VSD} the Victor-Purpura spike train distance $D_V$ is
plotted against the cost parameter $c_V$ for all pairs of spike trains in a group containing three members in
each of the three principal clusters. The relative order of the six different combinations of cluster
affiliations ($1-1$, $1-2$, $1-S$, $2-2$, $2-S$, $S-S$) depends on the cost parameter. At small $c_V$ values,
the Victor-Purpura distance measures the difference in spike counts and this number seems not to be closely
related to the type of underlying cluster (see in particular the spread of the $D_V$-values corresponding to the
$1-2$ combination). At high $c_V$ values, distances are larger and also very mixed. It is only at intermediate
values that a clear separation of the six different cluster combinations can be observed. High values are
obtained for all the $1-2$ combinations while the intra-cluster $S-S$ distance is the smallest one.

\begin{figure}
    \includegraphics[width=85mm]{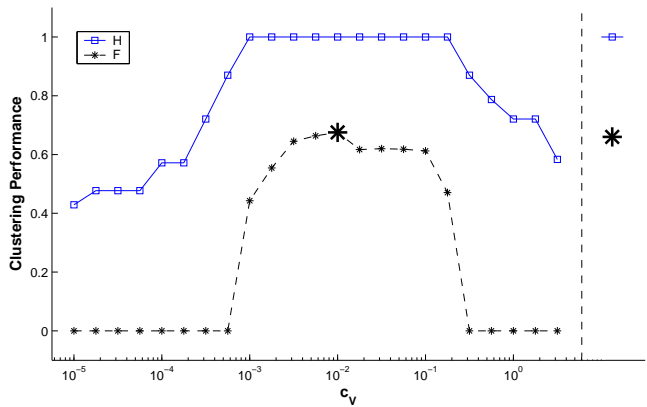}
    \caption{\abb\label{fig:Perf-VSD} For the Victor-Purpura spike train distance the clustering entropy $H$ (solid line)
        and the cluster separation $F$ (dashed line) in dependence of the parameter $c_V$ that sets the time
        scale. Remember that $F$ is only calculated for those cases where a correct clustering (as reflected by $H = 1$)
        is obtained. The optimum performance $F = 0.67$ is marked by a large asterisk at $c_V = 0.01$. On the
        right hand side of the dashed line the respective values for the ISI-distance ($H = 1$ and $F = 0.66$),
        which required no parameter optimization, are depicted.}
\end{figure}

We find similar results for the other parameter--dependent measures. In all cases, there exists an intermediate
parameter range where meaningful results can be obtained (i.e., the cluster entropy is equal to 1), while for
higher and lower values no clear clustering can be recognized. The role of the free parameter is better seen by
determining the clustering performance for different values of the parameter itself. In Fig.~\ref{fig:Perf-VSD}
the two indices $H$ and $F$ are plotted for the Victor-Purpura distance $D_V$ versus the cost parameter $c_V$.
We see that a perfect clustering is found only inside an intermediate range, where $H=1$; the smaller $H$ values
found outside this interval reflect the presence of misclassifications in the clustering tree. Inside the
interval where the right classification is obtained, we computed the clustering separation $F$. This index
attains its peak value $F = 0.67$ when $c_V = 0.01$, which we thus identify as the optimal value of the time
scale for the separation of the different clusters. This result is consistent with the visual impression when
looking at Fig.~\ref{fig:Para-VSD} (the vertical line corresponds to the optimal $c_V$ value). A similar
scenario is obtained also for the other measures that depend on a time scale. In each case there exists an
intermediate range where $H=1$. The broadest range is found for the Victor-Purpura and the van Rossum distance.

\begin{figure}
    \includegraphics[width=85mm]{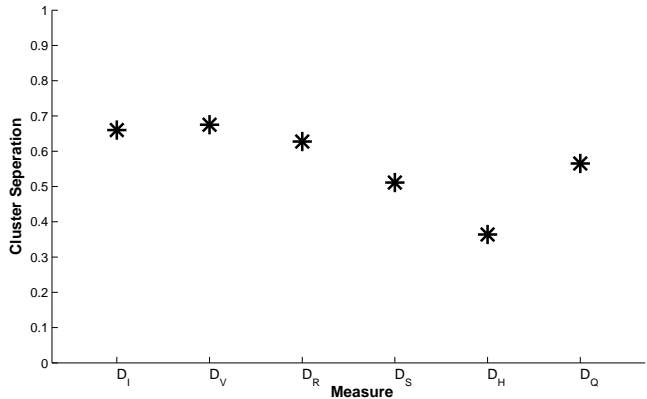}
    \caption{\abb\label{fig:Cluster-Separation} Comparison of Measures: Separation of clusters.}
\end{figure}

The performance of the different measures are compared in Fig.~\ref{fig:Cluster-Separation}, where the maximum
value of the cluster separation is shown for each spike train distance (for the parameter-free ISI-distance and
the event distance no optimization is required). The best results are found for the Victor-Purpura distance
$D_V$, but the ISI-distance $D_I$ and the van Rossum distance $D_R$ perform almost equally well. At the other
extreme, the poorest cluster separation is obtained for the Hunter-Milton dissimilarity $D_H$.

%

\subsection{\label{ss:Results-Correlation-among-measures} Correlations between the different measures}

In order to assess the degree of redundancy among the different measures of spike train dissimilarity, a
correlation and cluster analysis has been performed on all $406$ bivariate combinations of measure results by
following the approach discussed in the previous section. Since the number of independent observations can
hardly be estimated, this is only a relative examination. For this reason no values of significance are given.

\begin{figure}
    \includegraphics[width=85mm]{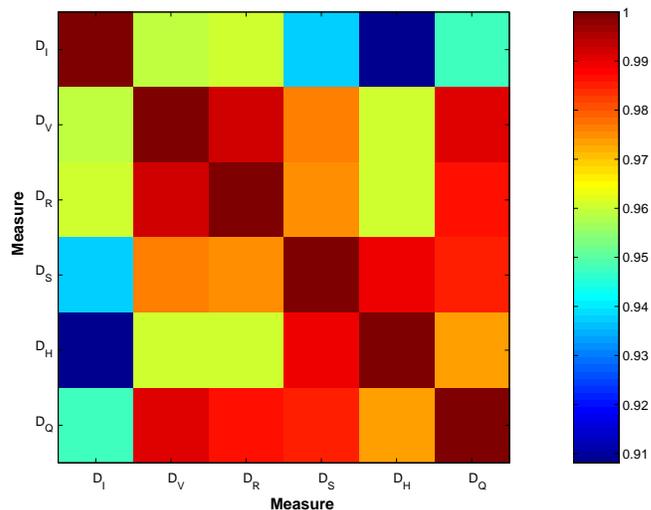}
    \caption{\abb \label{fig:MatCorr} (color online) Correlation coefficients between the six measures of spike train dissimilarity.}
\end{figure}

\begin{figure}
    \includegraphics[width=85mm]{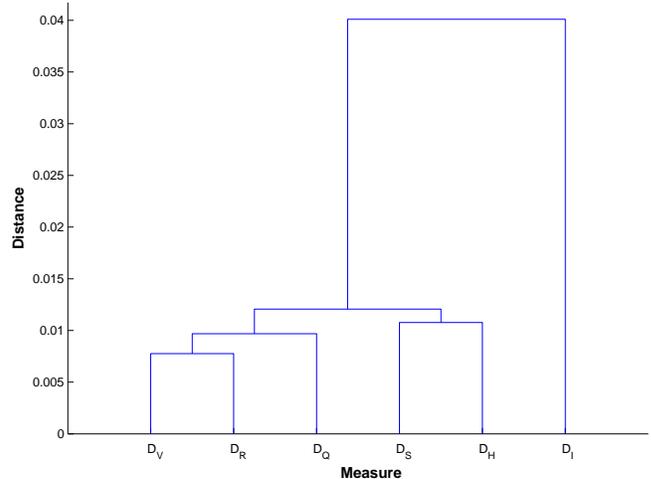}
    \caption{\abb\label{fig:MatCluster} Clustering for the six measures of spike train dissimilarity.}
\end{figure}

As we see from the high overall level of correlation shown in Fig. \ref{fig:MatCorr}, all measures seem to carry
similar information. The minimum correlation coefficient $0.91$ is obtained between the ISI-distance and the
Hunter-Milton dissimilarity, while the spike train distances by Victor-Purpura and van Rossum appeared to be the
most correlated measures. From the corresponding cluster tree (cf. Fig.~\ref{fig:MatCluster}), it becomes
evident that the ISI-distance is the most independent measure, whereas the other measures belong to one big
cluster. This reflects the fact that the ISI-distance is derived from interspike intervals, while the other
measures are based on spike times.

\section{\label{s:Discussion} Discussion}

In this study we propose a simple method for measuring the (dis-)similarity of two spike trains. As an estimator
based on the relative sizes of interspike intervals, the ISI-distance is complementary to all spike-based
measures of synchrony that quantify the simultaneous occurrences of spikes via some sort of coincidence
detection. As illustrated by an application to in vitro recordings of cortical cells, the measure serves also as
an excellent means to visualize the occurrence of spiking pattern in the respective spike trains. Finally, this
approach represents a natural starting point towards a more complete characterization of neuronal activity, in
so far as higher moments (such as the standard deviation) of the $I(t)$ distribution can be measured and
compared.

In order to judge the relative merit of the different methods, we compared the methods by evaluating each of
them on spike trains extracted from a network of simulated Hindemarsh-Rose neurons. We assessed the ability of
the different measures to reproduce the original clustering (established by a priori adjusting the synaptic
couplings in the model) by means of two indices. In this comparison, no measure fails in reproducing the
expected clustering; however, we found subtle differences in the degree of separation among the three clusters.
The ISI-distance performed as well as the best spike-based measure (the Victor-Purpura distance $D_V$) with the
distinct advantage of not requiring the optimization of any parameter. In fact, the ISI-distance, like event
synchronization, is self-adaptive in that it automatically identifies the proper time scale. In particular, this
holds true for changes of firing rate within the same spike trains. Whereas the ISI-distance can also be applied
to spike trains that include different time scales (i.e., regular spiking and bursting) the other measures would
misrepresent either behaviour depending on the parameter chosen.

Finally, we implemented a correlation analysis in order to evaluate the degree of independence among the six
different measures of spike train similarity. Since the overall level of correlation is quite high, all measures
apparently access similar information. The subsequent cluster analysis reveals that the ISI-distance is the most
independent approach. This is not surprising since the ISI-distance is the only measure that can be regarded as
a measure of rate coding (since it is built on instantaneous firing rate estimates) whereas all other measures
(that are based on spike timing) can be interpreted as measures of time coding.

Some general remarks concerning the application of measures of spike train (dis-)similarity: First, although the
focus of this study is on the estimation of similarity between just two spike trains, all methods can also be
used within a multivariate context. For example, in order to assess the reproducibility of neural spike train
responses to an identical stimulus across many different presentations (trials), reliability can be defined as
the average over all pairwise correlation-values (as done in \citeauthor{Haas02} (\citeyear{Haas02}),
\citeauthor{S_Schreiber03} (\citeyear{S_Schreiber03}), and \citeauthor{Hunter03} (\citeyear{Hunter03})). On the
other hand, it should be noted that the ISI-distance, like all the other measures, can be fooled by phase lags,
no matter whether these are caused by internal delay loops in one spike train or by a common driver that forces
the two time series with different delays. Thus any such phase lags should be removed by suitably shifting the
time series before computing the measures.

The comparisons carried out in this study constitute an important step towards the identification of the most
appropriate spike train (dis-)similarity for an application to real data. From the observed performance values
and the results of the correlation analysis, we conclude that the ISI-distance together with either one of the
spike based measures, such as the Victor-Purpura or the van Rossum spike train distance, or event
synchronization is the most appropriate choice. However, it should also be noted that other possible criteria
for the selection of an appropriate measure, such as computational cost or robustness against noise, remain to
be evaluated in future studies.

%
%
%
%

\begin{appendix} \label{s:Appendix}
%
%
\section{\label{s:Data} Data}

\subsection{\label{ss:Data-Stellate-Cells} Application to in vitro recordings of cortical cells}
The ISI-distance is illustrated using in vitro whole-cell recordings taken from cortical cells from the layer 2
medial entorhinal cortex of young Long-Evans rats. In these experiments (conducted as approved by the UCSD
IACUC) cortical cells were selected from a slice preparation ($400$ micron) by their superficial position, as
well as particular characteristics of their electrophysiological responses to long current steps (cf.
\citeauthor{Haas02}, \citeyear{Haas02}). Intracellular signals were amplified, low pass filtered, and digitized
at $10$ kHz via software created in LabView (National Instruments). Inputs were delivered as synaptic
conductances through a linux-based dynamic clamp (\citeauthor{Dorval01}, \citeyear{Dorval01}). Inputs were
comprised of synaptic inputs added to an underlying DC depolarization. The amplitude of the DC depolarization
was tailored for each cell to elicit a spike rate of $5-10$ Hz. Synaptic inputs were of the form $Isyn =
G_{syn}S(V_m-V_{syn})$ where $S$ follows the differential expression $dS/dt = \alpha (1-S) = \beta S$;
$\alpha=500/ms$, $\beta = 250/ms$, $V_{syn} = 0 mV$. $G_{syn}$ was tailored for each cell to be peri-threshold
and elicit a spike with probability close to $50\%$. Synaptic events were delivered for $10$ seconds at either
regular intervals or chaotic intervals; the latter intervals were taken from a set of five $60$ second
recordings of SC spike times in response to steady DC depolarization alone.

\subsection{\label{ss:Data-Hindemarsh-Rose} Hindemarsh-Rose simulations}
The spike trains have been generated using time series extracted from a larger network of Hindemarsh-Rose (HR)
model-neurons (\citeauthor{Hindmarsh84}, \citeyear{Hindmarsh84}) in the chaotic regime. This network was
originally designed to analyze semantic memory representations using feature-based models; details of the
network architecture and the implementation of the feature coding can be found in \citeauthor{Morelli06}
(\citeyear{Morelli06}).

In short, the state the neuron $i$ is determined by three first-order differential equations describing the
evolution of the membrane potential $X_{i}$, the recovery variable $Y_{i}$, and a slow adaptation current
$Z_{i}$,
\begin{eqnarray} \label{eq:Hindemarsh-Rose}
    \dot{X_{i}} &=& Y_{i}-X_{i}^{3}+3X_{i}^{2}-Z_{i}+I_{i}
        +\alpha_{i}(t)-\beta_{i}(t) \\
    \dot{Y_{i}} &=& 1-5X_{i}^{2}-Y_{i} \\
    \dot{Z_{i}} &=& 0.006[4(X_{i}-1.6)-Z_{i}],
\end{eqnarray}
where
\begin{equation} \label{eq:HR-impulse-current}
    \alpha_{i}(t) = \sum_{j=1}^{\widehat{F}(\widehat{M}-1)}w_{ij}A_{j}(t)
\end{equation}
and
\begin{equation} \label{eq:HR-local-inhibition}
    \beta_{i}(t) = \frac{1}{\widehat{F}-1}\sum_{k=1}^{\widehat{F}-1}A_{k}^{(i)}(t).
\end{equation}

The network consisted of $\widehat{N}=128$ HR neurons belonging to $\widehat{M}=16$ different modules with
$\widehat{F}=8$ neurons each. In a learning stage, input memory patterns were stored by updating the synaptic
connection weights $w_{ij}$ between different neurons using a Hebbian mechanism based on the activity variables
$A_j$. During the retrieval stage in which the learned connection weights were kept constant, the $29$ time
series to be analyzed were extracted. According to their coding properties regarding the retrieval of two
distinguished memory patterns, the $29$ time series belonged to three principal clusters, $13$ of the
corresponding neurons coded for pattern $1$ only, $13$ coded for pattern $2$ only and $3$ coded for both
patterns (shared). The respective time series were labelled by '1', '2' and 'S' followed by an index letter. The
numerical integration was done by using a fixed-step fourth-order Runge-Kutta method. The integration step-size
was chosen equal to $0.05$ ms of real time. The length of the time series analyzed was $32768$ data points.

\end{appendix}

\vspace{1cm}

\begin{thanks}
\section{\label{s:Acknowledgement} \textbf{Acknowledgements}}
We thank S. Luccioli, A. Torcini and K. Ulbrich for useful discussions and P. Grassberger for carefully reading
this manuscript. TK has been supported by the Marie Curie Individual Intra-European Fellowship "DEAN", project
No 011434. JSH acknowledges financial support by the San Diego Foundation.
\end{thanks}


\bibliographystyle{unsrtnat}

\end{document}